\documentclass[conference]{IEEEtran}
\IEEEoverridecommandlockouts
\usepackage{cite}
\usepackage{amsmath,amssymb,amsfonts}
\usepackage{algorithmic}
\usepackage{graphicx}
\usepackage{textcomp}
\usepackage{xcolor}
\usepackage{threeparttable}
\usepackage{mathtools}
\usepackage{xspace}
\usepackage{graphicx}
\usepackage{wrapfig}
\usepackage{lscape}
\usepackage{rotating}
\usepackage{epstopdf}
\usepackage{siunitx}
\usepackage{caption}
\usepackage{subcaption}
\usepackage{xspace} 
\usepackage{lipsum}
\usepackage[utf8]{inputenc}
\usepackage{mathtools}
\usepackage{xcolor}
\usepackage{soul}

\DeclarePairedDelimiter{\abs}{\lvert}{\rvert} 

\usepackage{fancyhdr}   

\fancypagestyle{firstpage}{     
  \fancyhf{}
  \chead{\textcolor{gray}{This article has been accepted for publication in the proceedings of the \\ 7th International Conference on Activity and Behavior Computing}}
  \fancyfoot[C]{\small{\textcolor{gray}{~\copyright~ 2025 IEEE.  Personal use of this material is permitted.  Permission from IEEE must be obtained for all other uses, in any current or future media, including reprinting/republishing this material for advertising or promotional purposes, creating new collective works, for resale or redistribution to servers or lists, or reuse of any copyrighted component of this work in other works.}}}

}

\def\BibTeX{{\rm B\kern-.05em{\sc i\kern-.025em b}\kern-.08em
    T\kern-.1667em\lower.7ex\hbox{E}\kern-.125emX}}
\begin{document}

\title{Hybrid CNN-Dilated Self-attention Model Using Inertial and Body-Area Electrostatic Sensing for Gym Workout Recognition, Counting, and User Authentification 
}

\author{\IEEEauthorblockN{Sizhen Bian}
\IEEEauthorblockA{\textit{DFKI} \\
Kaiserslautern, Germany \\
sizhen.bian@dfki.de}
\and
\IEEEauthorblockN{Vitor Fortes Rey}
\IEEEauthorblockA{\textit{DFKI} \\
Kaiserslautern, Germany \\
vitor.fortes\_rey@dfki.de}
\and
\IEEEauthorblockN{Siyu Yuan}
\IEEEauthorblockA{\textit{RPTU} \\
Kaiserslautern, Germany \\
syuan@rptu.de}
\and
\IEEEauthorblockN{Paul Lukowicz}
\IEEEauthorblockA{\textit{DFKI} \\
Kaiserslautern, Germany \\
paul.lukowicz@dfki.de}
}

\maketitle
\begin{abstract}

While human body capacitance ($HBC$) has been explored as a novel wearable motion sensing modality,  its competence has never been quantitatively demonstrated compared to that of the dominant inertial measurement unit ($IMU$) in practical scenarios.  This work is thus motivated to evaluate the contribution of $HBC$ in wearable motion sensing. A real-life case study, gym workout tracking, is described to assess the effectiveness of $HBC$ as a complement to $IMU$ in activity recognition. Fifty gym sessions from ten volunteers were collected, bringing a fifty-hour annotated $IMU$ and $HBC$ dataset. With a hybrid CNN-Dilated neural network model empowered with the self-attention mechanism, $HBC$ slightly improves accuracy to the $IMU$ for workout recognition and has substantial advantages over $IMU$ for repetition counting. This work helps to enhance the understanding of $HBC$, a novel wearable motion-sensing modality based on the body-area electrostatic field. All materials presented in this work are open-sourced to promote further study \footnote{https://github.com/zhaxidele/Toolkit-for-HBC-sensing}.


\end{abstract}


\section{Introduction}

\thispagestyle{firstpage} 

Human Activity Recognition (HAR) deciphers human actions through advanced sensing and computing techniques to enable the machine to comprehend, analyze, understand these actions and bring assisted services correspondingly \cite{yan2022emoglass, bian2020wearable}.
Advances in wearable sensor technique have enabled a broader range of HAR applications, like interaction \cite{bian2021capacitive, azim2022over}, gaming\cite{ lai2019fitbird},  clinic\cite{parker2018interplay, garcia2024challenges, fikry2024improving, garcia2024relabeling},  healthcare\cite{nishimura2022toward,bian2019passive, kuosmanen2022does, tazin2024understanding, victorino2024forecasting}, sport\cite{tanaka2023full}, etc. 
Among the rich set of wearable sensors, the inertial measurement unit ($IMU$) plays a substantial role for wearables that provide motion knowledge of individuals \cite{xu2021hulamove, bian2022state}. 
Although $IMU$ has been regarded as the "gold standard" solution for wearable motion sensing, it faces several intrinsic challenges that are impossible to address by itself, such as the accumulated error issue resulting from continually integrating acceleration with respect to time to calculate velocity or position (for tracking scenarios); and the locality limitation, which means that the $IMU$ only perceive motion status of where it is deployed. In wearables, motion information from remote body parts often gets damped or lost when reaching the sensor deployment position, the inaccuracy of $IMU$-based step counting on wrist-worn devices as an example \cite{toth2019effects, bian2024earable}. To enhance wearable motion sensing, especially for HAR tasks, extended sensing functionality by sensor fusion is the most straightforward yet effective way. However, an $IMU$-alternative/complementary, cost- and power-efficient sensing modality doesn't exist in commercial wearable devices yet.

This paper evaluates the effective contribution of a potential wearable motion sensing modality that is still not fully explored but shows impressive potential in wearable HAR: the passive body-area electrostatic field \cite{bian2024body}, also indicated as human body capacitance ($HBC$).
An easily overlooked fact is that the human body, the subject for wearable body motion tracking tasks, has good intrinsic electric features due to its ideal conductivity \cite{presta1983measurement,cochran1986total}.  
Early studies \cite{aliau2012fast, aliau2013novel, buller2006measurement, greason1995quasi} have demonstrated (mathematically or physically) that the value of body capacitance is a dynamic one varying from 100$pF$ to 400$pF$. However, the aim of early $HBC$ studies was mostly to develop shielding techniques of electrostatic discharge for safety operations. The idea of making use of $HBC$ only emerged gradually from the last decades, examples are $HBC$-based communication \cite{cohn2012humantenna}, cooperation detection \cite{bian2019wrist} as well as basic motion detection \cite{cohn2012ultra, bian2022using}. 
Despite such pioneer studies, the question of how this $HBC$ concept could benefit current wearable motion sensing remains unanswered. Thus, to further explore the potential of $HBC$ (as the only one that is efficiency-competitive in cost and power consumption with the dominant $IMU$) for wearable motion sensing, more quantitative and comparable studies in practical scenarios are required.
A previous work \cite{bian2019passive} has preliminarily demonstrated the feasibility of $HBC$-based activity recognition in fitness (with an accuracy of 63\% of seven workouts under leave-one-user-out evaluation). To evaluate the effective contribution of $HBC$ in HAR, especially when compared with the dominant $IMU$-based solution, this work analyzes the performance of workout tracking with the signal source of $HBC$ and $IMU$ separately and jointly, aiming to provide a first rigorous conclusion on the benefaction of $HBC$ to existing wearable motion sensing solutions.  
As a popular topic, fitness tracking research appeared in a rich set of literary works, either mobile phones or stand-alone $IMUs$ are used to detect the object's movements. For example, the authors in \cite{koskimaki2014recognizing} explored 30 exercises and got an outstanding recognition result. However, since the data was from only one subject, the evaluation lacks generalization. In \cite{wahjudi2019imu}, the authors deployed the $IMU$ on shoes and analyzed the gait to recognize walking-related workouts. \cite{depari2019lightweight} focused on the free-weight exercises and reported 93\% accuracy with a single $IMU$. Most of such works' signal source is the $IMU$ alone. 
In this work, we collected a workout dataset composed of twelve gym activities with both $IMU$ and $HBC$ sensing. By classifying the workouts with an attention-based deep neural network model,
we concluded that the $HBC$ could be a valuable supplementary signal for body motion-related research and applications. 

\section{Related works of wearable motion sensing modalities}

Wearable sensors have been widely explored in the past decades for lifestyle characterization and health monitoring, and a large part of them have been evaluated particularly for detecting human body motion. 
The inertial sensor is undoubtedly the most popular modality as it is widely integrated into everyday smart devices, benefiting from low cost and power consumption. However, the inertial sensor is designed to sense the motion pattern of the object where it is attached. In human activity recognition, the actions from the non-attached body part (e.g., legs) are intuitively not able to be sensed by a remote inertial sensor, which is attached to another body part (e.g., wrist), or suffering motion knowledge loss during the action transmission (from legs to wrist). Vision sensors, like the camera, are commonly explored in the computer vision domain for robotics applications. Yet, researchers also studied the wearable-form camera for activity recognition, such as \cite{cartas2020activities}, where the authors deployed the camera in front of the user's chest for egocentric activity recognition. Despite the fruitful context captured by cameras and advanced algorithms dealing with image data, cameras in wearable form suffer from uncomfortable wearing for users, and the signal processing also suffers from high computing load. Pressure sensors in wearable form, normally in the form of the insole, for human activity recognition is also a widely explored study. Such sensing prototypes often deliver high accuracy for the target application due to the grid-wise sensing node distribution, but the cost of building such a sensing system makes it fall behind the other wearable sensors that only need a few dollars \cite{dong2021design}. The magnetic field sensing system is robust to everyday environments and brings especially the absolute accurate position information that other wearable sensors are incapable \cite{bian2020wearable, bian2020social}. However, such systems rely on the coils to generate and sense the magnetic field, which results in a large hardware volume and is uncomfortable to wear. Recent studies on inaudible acoustic signals show impressive deliveries in daily activity recognition \cite{mahmud2024actsonic}, and the sensing system can be miniaturized into a glass frame and deployed on the eyewear. However, such acoustic signals for activity recognition raise concerns about environmental audio dependency. Besides the above sensing modalities, RF signals, such as Bluetooth signals, were also studied for motion sensing, especially in position-related explorations. Wearable Bluetooth devices enjoy the advantages of low power, low cost, and compact form factor, but the limitation is also evident: the signal is actively generated, consuming more power than passive sensing modalities, and a server around it is commonly needed for signal strength perception. As can be seen, an $IMU$-competitive wearable motion sensing modality regarding the hardware size, cost, computing load, and power consumption, which are critical features for practical deployment, has so far not been presented. Such a background has motivated us to conduct a comparative investigation of the body-area electrostatic field sensing modality triggered by its characteristics in cost, volume, and power consumption, as well as the potential for practical deployment.

\begin{table*}[]
\centering
\begin{threeparttable}

\begin{tabular}{ p{0.6cm} p{0.7cm} p{0.7cm} p{0.7cm} p{1.0cm} p{1.0 cm} p{3.2cm} p{3.3cm} p{1.2cm}}
& size & cost & power & linearity & output & circuit form & attach position on body & robustness \\ 
\hline
$IMU$ & $mm^2$  & dollars & $uW$ & yes & digital & integrated circuit chip & motion part & strong \\
\hline
$HBC$ & $mm^2$ & cents & $uW$ & no & analog & discrete components & motion/static part & weak \\ 
\hline
\end{tabular}
\caption{$IMU$ and $HBC$ sensing front-end comparison}
\label{Comparision}
\end{threeparttable}
\end{table*}

\begin{figure}[]
\centering
\begin{minipage}[t]{0.45\linewidth}
\centering
\includegraphics[width=0.99\textwidth,height=2.5cm]{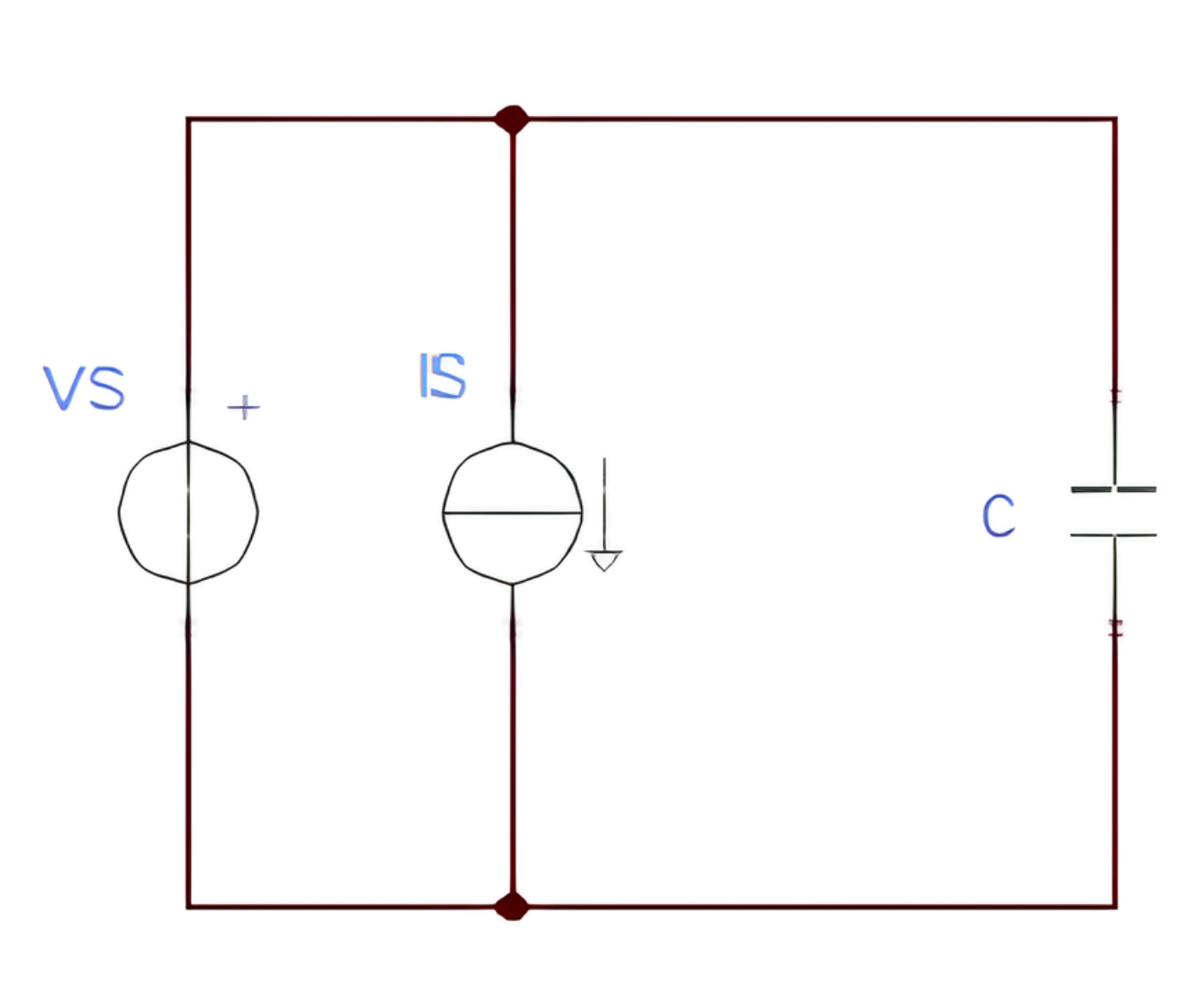}
\caption{Basic structure of a body capacitance sensing method}
\label{Sensor}
\end{minipage}
\centering
\quad
\begin{minipage}[t]{0.45\linewidth}
\centering
\includegraphics[width=0.8\textwidth,height=2.5cm]{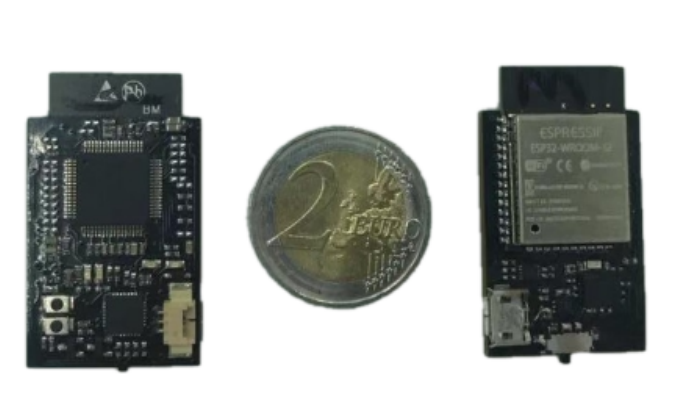}
\caption{The prototype for body motion sensing including $IMU$ and $HBC$ sensing}
\label{Prototype}
\end{minipage}
\centering
\end{figure}

\section{Physical Background and Sensing Prototype}
Instead of monitoring the body capacitance directly, we continuously measured the potential variation from the body surface. Variation in $C_B$ will result in the charge flow on the surface of the human body. Figure \ref{Sensor} shows the basic principle of our sensing design where $C$ refers to the body-area electrostatic field. The voltage source maintains the potential level of the body surface-coupled or -contacted electrode; the current source supplies electrons to $C$. Once $C$ varies, a potential variation will occur immediately. Then the potential level returns to the level of $VS$ with the complement of the electrons from $IS$. The whole mechanics is a series of charging and discharging processes. 
Figure \ref{Prototype} shows our prototype for body motion tracking, which comprised a battery charging module, an ESP32 processing unit with WIFI and BLE integrated, a 24-bit high-resolution ADC ads1298, an $IMU$, and a $HBC$ sensing front-end composed of several discrete components. The high-cost ADC can be replaced by a simple amplification circuit to maintain the low cost of the system. 
The discrete components, which are the core of the sensing front-end, include two nF-level capacitors and four up to Mohm-level resistors, consuming only $uW$ level of power. The schematic of the system is open-sourced for reproduction and further development.
We used the standard 43mm ECG electrode as a connection medium between our prototype and the human body surface. 
Table \ref{Comparision} summarizes a shallow comparison between $IMU$ and $HBC$ sensing front end. The $HBC$ sensing modality enjoys similar advantages to $IMU$ in size, cost, and power consumption. It outperforms with the cross-body part attachment feature due to the body’s conductivity. The limitation of $HBC$ sensing is its non-linear output and sensitivity to surroundings (weak robustness). 
In the following sections, the contribution of each sensing modality to motion detection, with the gym workout recognition and counting as the case study, is evaluated.

\begin{figure}[]
\centering
\includegraphics[width=0.99\linewidth,height=8cm]{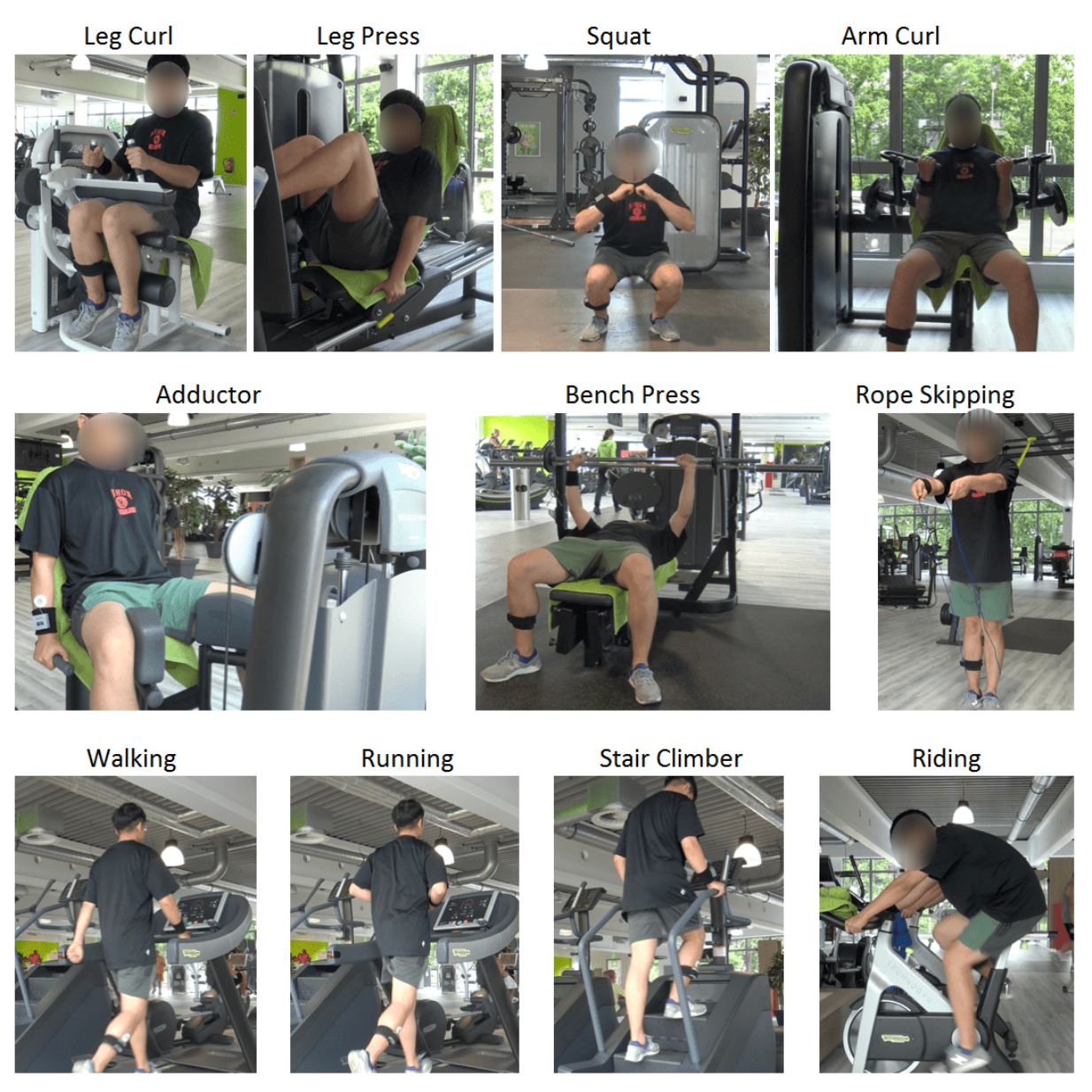}
\caption{Eleven gym workouts}
\label{Eleven}
\end{figure}

\begin{figure*}
\centering
\includegraphics[width=0.99\linewidth,height=11.5cm]{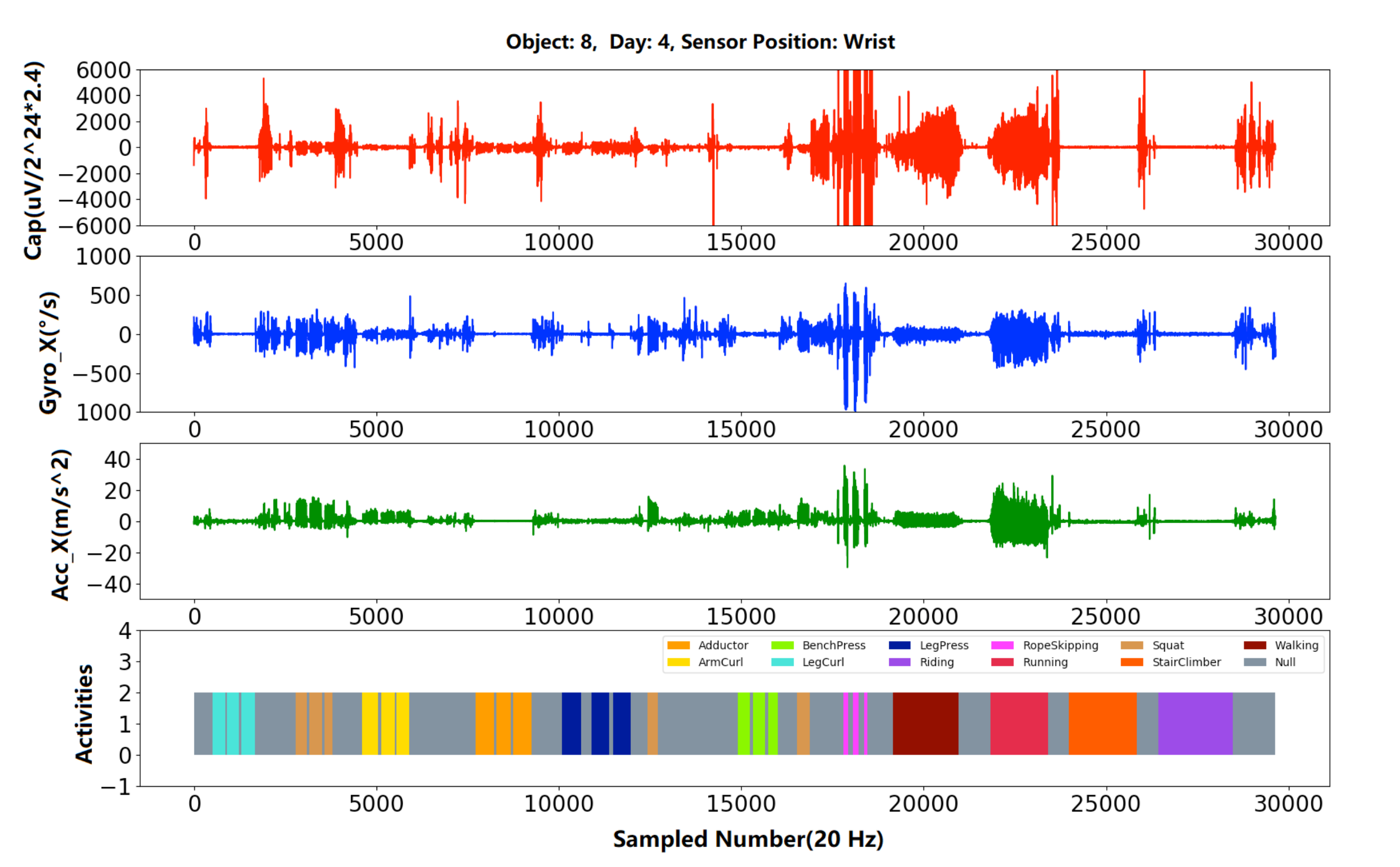}
\caption{Example of one session's initial measurement unit signal, capacitance caused potential variation signal, and the exercise labels, including a null class}
\label{Session}
\end{figure*}

\section{Contribution Evaluation}
\subsection{Data Collection}
We chose eleven popular gym exercises for the recognition task, including both aerobic and anaerobic training: Adductor, Armcurl, Benchpress, Legcurl, Legpress, Riding, Ropeskipping, Running, Squat, Stairsclimber, and Walking, as Figure \ref{Eleven} depicts. Both muscle strength and muscle endurance are trained. All the core muscle groups are considered within those eleven exercises, including pecs, quads, calves, biceps, triceps, gluteus, and hamstring. Running and Walking were performed on the treadmills with the speed of 5$\pm$0.2 km/h and 8$\pm$0.5 km/h for around 2 minutes each session. Riding and Stairclimbing were done at a self-determined speed and lasted around 2 minutes for each session. The rest of the exercises were trained with gym instruments(except Squat) for 3$x$10 repetitions each session. Ten subjects participated in this study, including five females and five males, with age from 21 to 30, weight from 49kg to 85kg, and height from 158cm to 184cm. Eight of them go to the gym at least three times a week, and two of them are novices. Each participant performed the above-listed exercises in five days. During the whole data collection phase, the temperature ranged from \SI{17.0}{\celsius} to \SI{27.5}{\celsius}, the relative humidity ranged from 45\% to 79\%(Data was from WetterKontor GmbH, measured by HMP45D). Since the triboelectric effect \cite{castle1997contact} will influence the electron distribution on the body, we also considered the wearing of each subject: height of shoe sole, sole shoe material (PVC or rubber), clothes material (polyester or cotton). Table \ref{Subjects_Configuration} shows the configuration of the participants' wearing. The aim of this configuration is to demonstrate the robustness of $ HBC$-based sensing modality in gym activity recording regarding wearing. To be noticed, the demonstration of $HBC$ robustness specified here only refers to the gym activity recognition and counting result (especially regarding the wearings), as we will apply leave-one-user-out cross-validation for the assessment, indicating the robustness of our classification and counting results. The "weak" robustness stated in Table 1 infers to the signal itself (not related to a specified $HBC$-based application result). 
The prototype was tested with three deployment positions: on the wrist, on the leg, and in the pocket.  

\begin{table}[]
\centering
\begin{threeparttable}
\begin{tabular}{ p{2.2cm} p{0.8cm} p{0.8cm} p{0.8cm} p{0.8cm} p{0.8cm}}
& First Day & Second Day & Third Day & Fourth Day & Fifth Day \\ 
\hline
Clothes Material & cotton & cotton & polyester & cotton & cotton\\
\hline
shoe sole height\tnote{a} & M & M & M & S & M  \\ 
\hline
shoe sole material & PVC & PVC & PVC & PVC & rubber  \\ 
\hline
\end{tabular}
\begin{tablenotes}
    \item[a] For each user, M, and S denote the height of the shoe sole, with M meaning the height of the pair of shoes the user is used to wearing, while S denotes the different height of another shoe belonging to the subject.
\end{tablenotes}
\caption{Participants' wearing configuration across days}
\label{Subjects_Configuration}
\end{threeparttable}
\end{table}

We collected the data with a frequency of 20 Hz and developed a user interface tool to get data from the prototype's Bluetooth. The data then got stored and displayed locally on a computer. During the gym training, a second person labeled all the exercises with the help of the developed tool by simply choosing and clicking. Overall, we got five sessions of a whole day's training for each subject with each sensor position. Each session involves around one hour's $IMU$ data and one hour's $HBC$ related body potential data. Within each session, there are three segments of each exercise(Adductor, Armcurl, Benchpress, Legcurl, Legpress, Ropeskipping, Squat), and one segment of each exercise(Riding, Running, Stairsclimber, Walking). Figure \ref{Session} depicts one whole session of $IMU$ and $HBC$ related potential data from the eighth subject on the fourth training day with the prototype worn on the wrist. 
To be noticed, we did another two kinds of Squat in each session, as depicted with the color Peru in Figure \ref{Session}. 
Each subject did the Squat with three ground types: concrete, wood, and rubber. The purpose was to verify the robustness of $ HBC$-based sensing modality related to different gym ground types, and the result is described in the evaluation result section.


In the beginning, we used the sliding window approach to get instances. The window size also plays a role in segmentation and classification \cite{banos2014window}, which is mainly a trade-off between the recognition speed and accuracy. We tried window size with 2, 4, and 6 seconds. The 4-second size performs best in our case, meaning each instance owes 80 readings. The overlapping size is 2 seconds.
To classify the whole session's activity, we defined another class named "Null", indicating the process when the trainee was not busy with the above-listed exercises. This class was depicted with the color grey in Figure \ref{Session}. 




\begin{figure}
\centering
\includegraphics[width=0.99\linewidth,height=7.0cm]{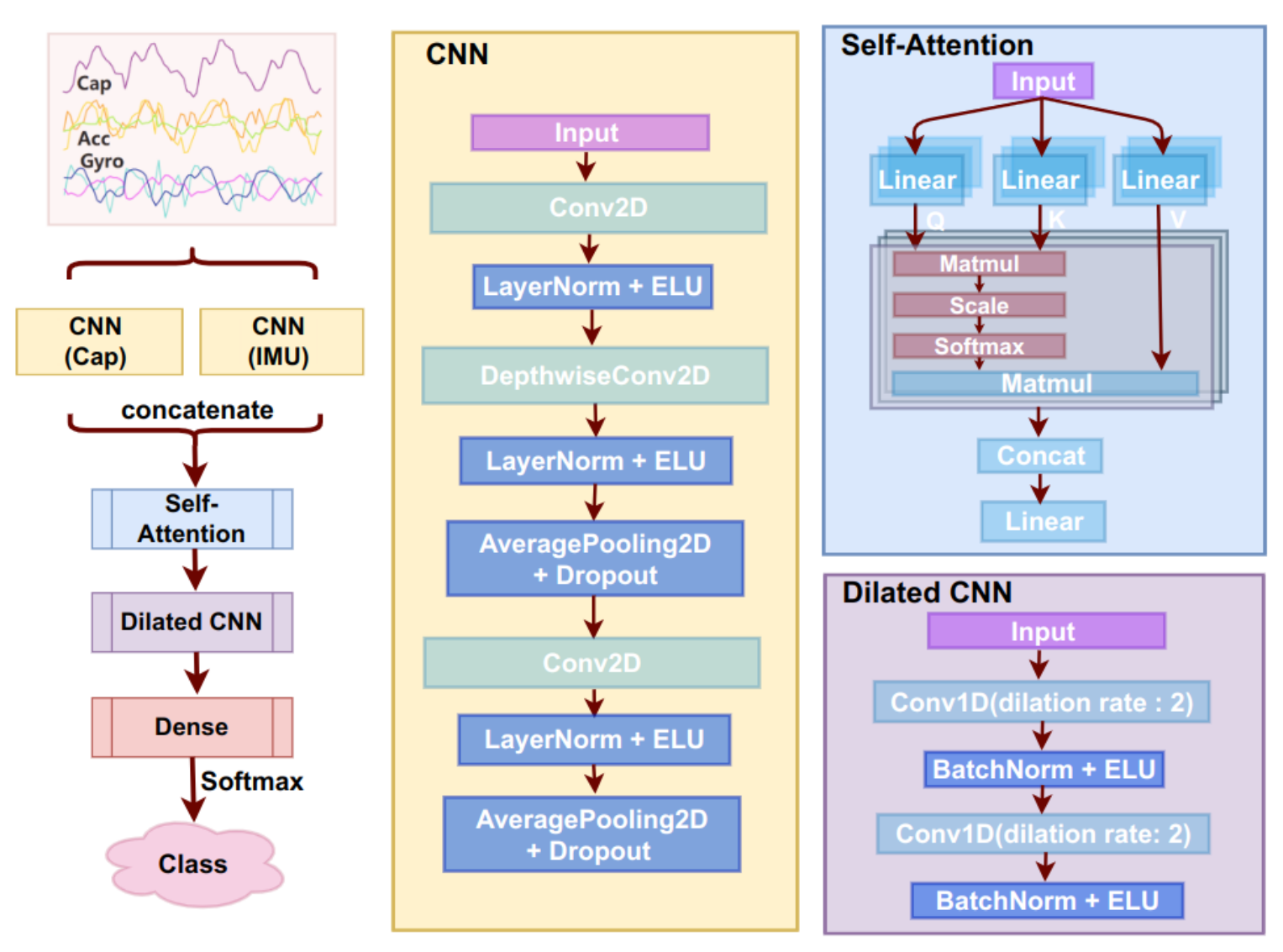}
\caption{Hybrid CNN-Dilated model embedded with the self-attention mechanism for workout recognition}
\label{Network}
\end{figure}

\begin{figure*}[!t]
\centering
\includegraphics[width=1.0\textwidth,height=13cm]{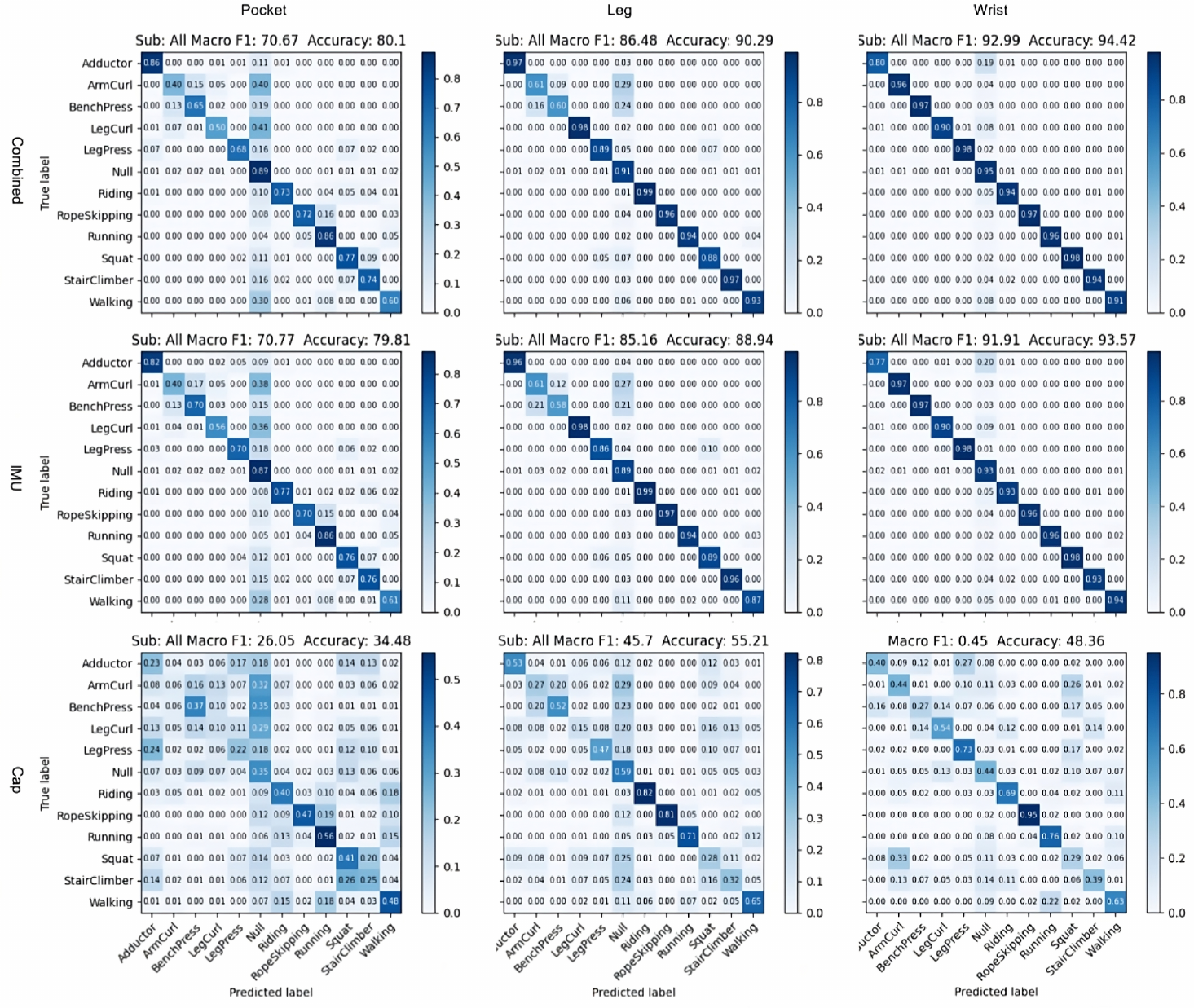}
\caption{Confusion matrix of classification result from the hybrid model}
\label{Attention_CM}
\end{figure*}


\begin{table*}[htbp]
\centering
\begin{threeparttable}
\caption{Classification of deep models: F-score/Accuracy}
\label{Deep_Resualt}
\begin{tabular}{p{2.4cm} p{2.0cm} p{2.4cm} p{2.4cm} p{2.6cm}}
\hline
Model & Position & $HBC$ & IMU & $HBC$+IMU \\ 
\hline

&Pocket & 0.25 / 0.31 & 0.67 / 0.78 & 0.66 / 0.78 \\
RF&Leg & 0.45 / 0.54 & 0.82 / 0.88 & 0.83 / 0.88 \\ 
&Wrist & 0.41 / 0.50 & 0.90 / 0.92 & 0.91 / 0.93\\ 
\hline

&Pocket & 0.17 / 0.16 & 0.53 / 0.62 & 0.57 / 0.62 \\
DeepConvLSTM&Leg & 0.23 / 0.21 & 0.62 / 0.69 & 0.69 / 0.71 \\ 
&Wrist & 0.17 / 0.16 & 0.59 / 0.66 & 0.63 / 0.66\\ 
\hline

&Pocket & 0.15 / 0.17 & 0.55 / 0.60 & 0.60 / 0.63 \\
Resnet21 & Leg & 0.39 / 0.41 & 0.78 / 0.81 & 0.76 / 0.75 \\
&Wrist & 0.32 / 0.37 & 0.89 / 0.91 & 0.89 / 0.91\\ 

\hline

& \textbf{Pocket} & \textbf{0.260 / 0.345} & \textbf{0.708 / 0.798} & \textbf{0.707 / 0.801} \\
\textbf{This work} & \textbf{Leg} & \textbf{0.457 / 0.552} & \textbf{0.852 / 0.889} & \textbf{0.865 / 0.903} \\
& \textbf{Wrist} & \textbf{0.450 / 0.484} & \textbf{0.919 / 0.936} & \textbf{0.930 / 0.944}\\

\hline
\end{tabular}
\end{threeparttable}
\end{table*}

\subsection{Classification Exploration}

Over the past decade, researchers have presented a rich set of algorithms for activity recognition tasks \cite{singh2023recent, yadav2021review, kumar2024human}. Both classic machine learning approaches, like SVM and RF, and deep neural network approaches, like CNN and Transformer, were presented and showed outperforming recognition performance compared to others in their specific activity recognition tasks. In this work, we designed a hybrid CNN-Dilated network empowered by the self-attention mechanism for workout recognition, achieving the best accuracy among tested models, as Figure \ref{Network} depicts. The network comprises three main blocks, a CNN block to extract spatial and temporal features from the raw signal, a multi-head self-attention block to bias the selection of most related values, and a dilated convolutional block to learn relations in a more extended sequence.

In the CNN block, a 2D convolutional layer with 32 kernels of size (1, SamplingRate/2) is first applied to extract features in the frequency domain from each channel separately. The second layer applied a depthwise convolution with the kernel size of (channel, 1) to extract spatial features from the frequency feature maps. Two filters are applied to each channel. The third convolutional layer consists of 128 filters of size (1, 10) again to strengthen the feature extraction in the temporal space. Each CNN layer is followed by a layer normalization to stabilize the training and an ELU activation to introduce nonlinearity. Average pooling (1, 2) and dropout (0.1) layers are applied to the second and third layers to downsample the feature maps and avoid overfitting. The output sequence of the CNN block is then split into four windows, and each is applied with a four-head self-attention model and dilated convolution. The self-attention blocks use the dot-product-based attention score to transform the feature values of each time step, which brings varying attention across the temporal domain. The weighted feature values are then fed into a two-layer dilated convolution structure (32 filters, each with kernel size 3) with a dilation rate of two for each to extract high-level features and generate the final representation, which is then used by a dense layer with softmax activation as the classifier. 
The network is trained with the categorical cross-entropy loss function and the Adam optimizer, with a 0.0001 initial learning rate and 0.5 decay rate every 200 decay steps. Ten-fold cross-validation is scheduled for training, where the leave-one-user-out strategy is applied. Each fold is trained 1000 epochs with early stopping using patience of 100. To balance the labels (obviously more "Null" samples among the instances), we apply the sample-weight method during training, and the corresponding weight of each sample is decided by the instance number of each class. The batch size is set to 256, and the training framework is Tensorflow-backend Keras.
Three signal sources are tested: the $HBC$ alone, the $IMU$ alone, and the combination. In the combination case, the spatial and temporal features of each sensing modality from the CNN block are concatenated before the multi-head self-attention block; as such post-fusing showed slightly better results than fusing at the input layer \cite{suh2023worker}.
Hyperparameter tuning relies more on experimental results than theory. Thus we determined the optimal settings by trying different combinations and evaluating the performance of each configuration, mainly involving the initial learning rate, kernel size and number in each layer, as well as the activations. Training was performed locally on a desktop with a NVIDIA GeForce RTX 4070 GPU.

\begin{figure*}[]
\centering
\includegraphics[width=1.0\textwidth,height = 5.0cm]{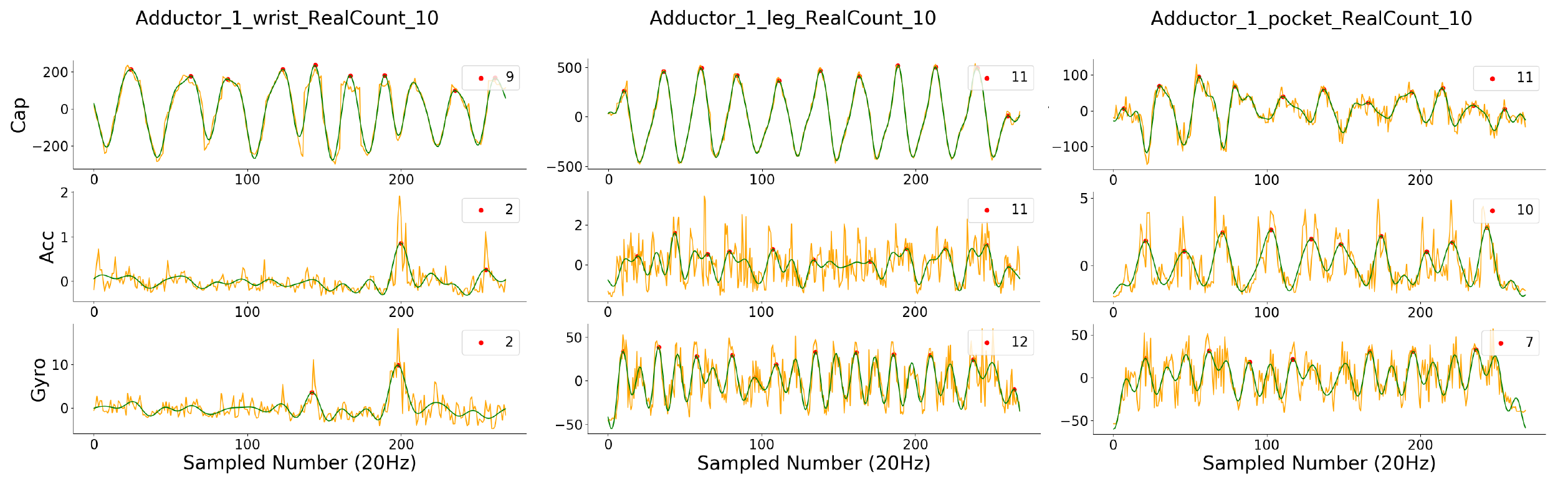}
\caption{Counting examples of Adductor workout}
\label{Counting_examples}
\end{figure*}

\begin{figure}[]
\begin{subfigure}{0.99\linewidth}
\centering
\includegraphics[width=8cm,height=5.0cm]{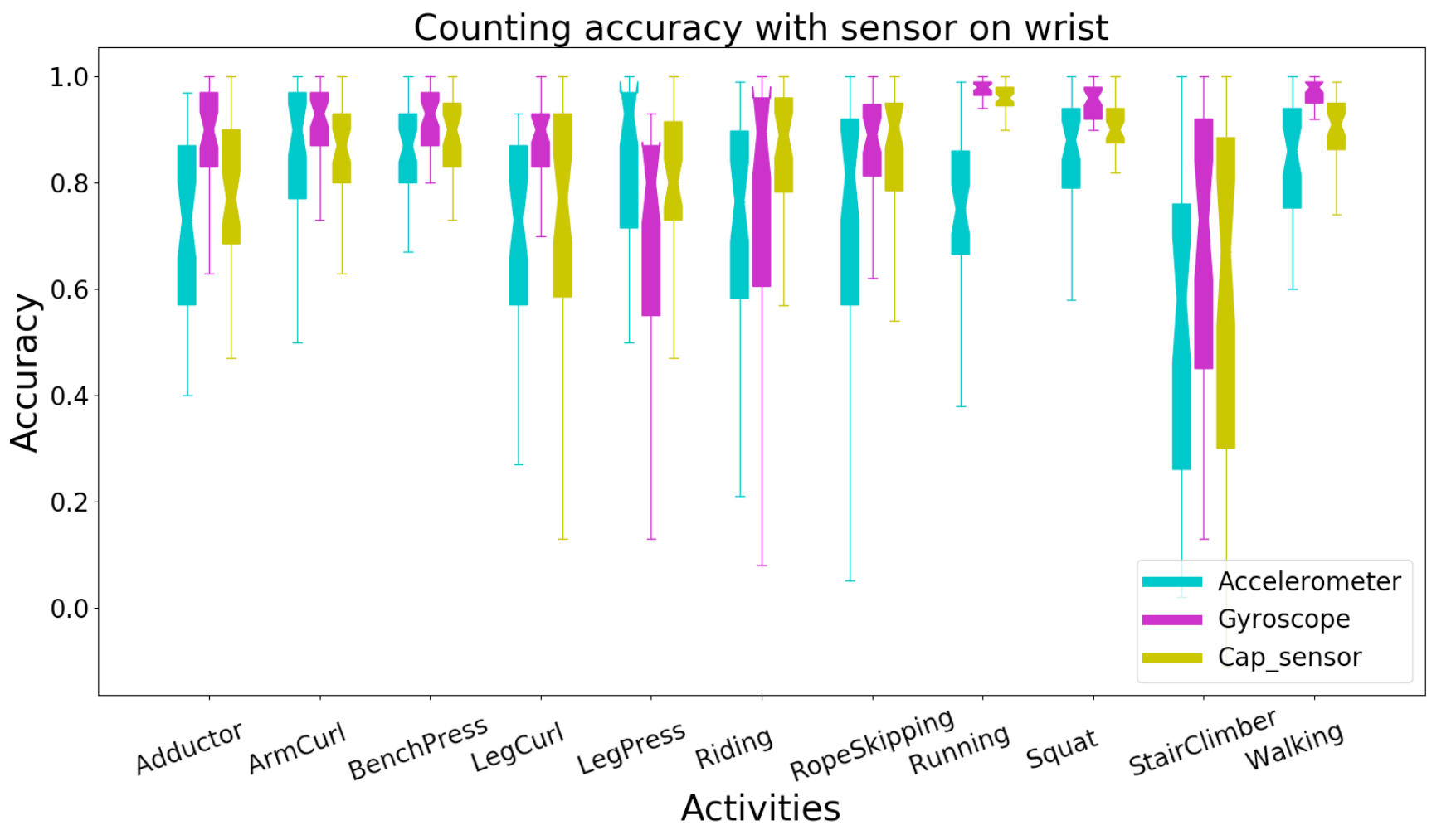}
\label{accuracy_wrist}
\end{subfigure}%
\\
\begin{subfigure}{0.99\linewidth}
\centering
\includegraphics[width=8cm,height=5.0cm]{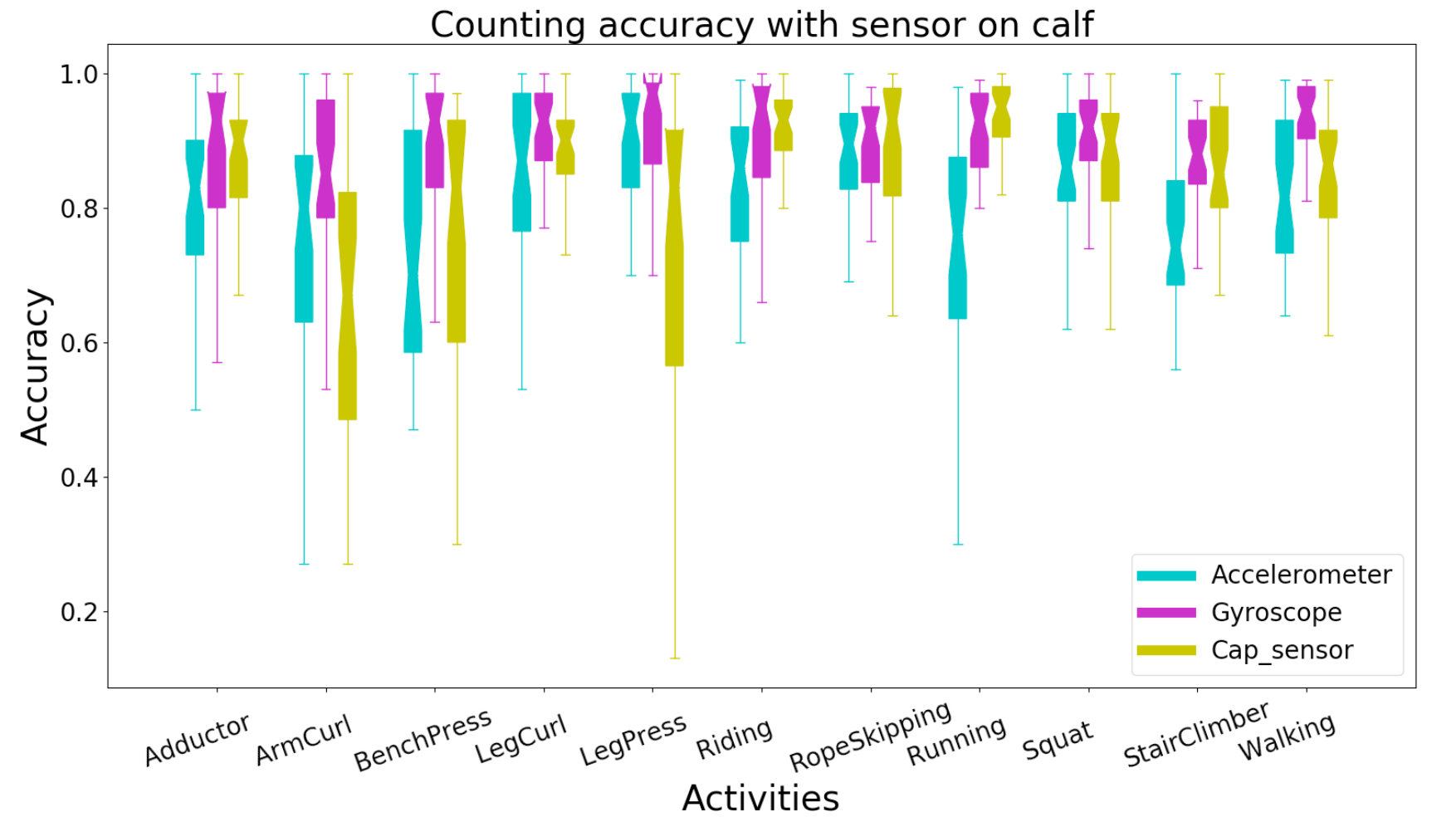}
\label{accuracy_leg}
\end{subfigure}%
\\
\begin{subfigure}{0.99\linewidth}
\centering
\includegraphics[width=8cm,height=5.0cm]{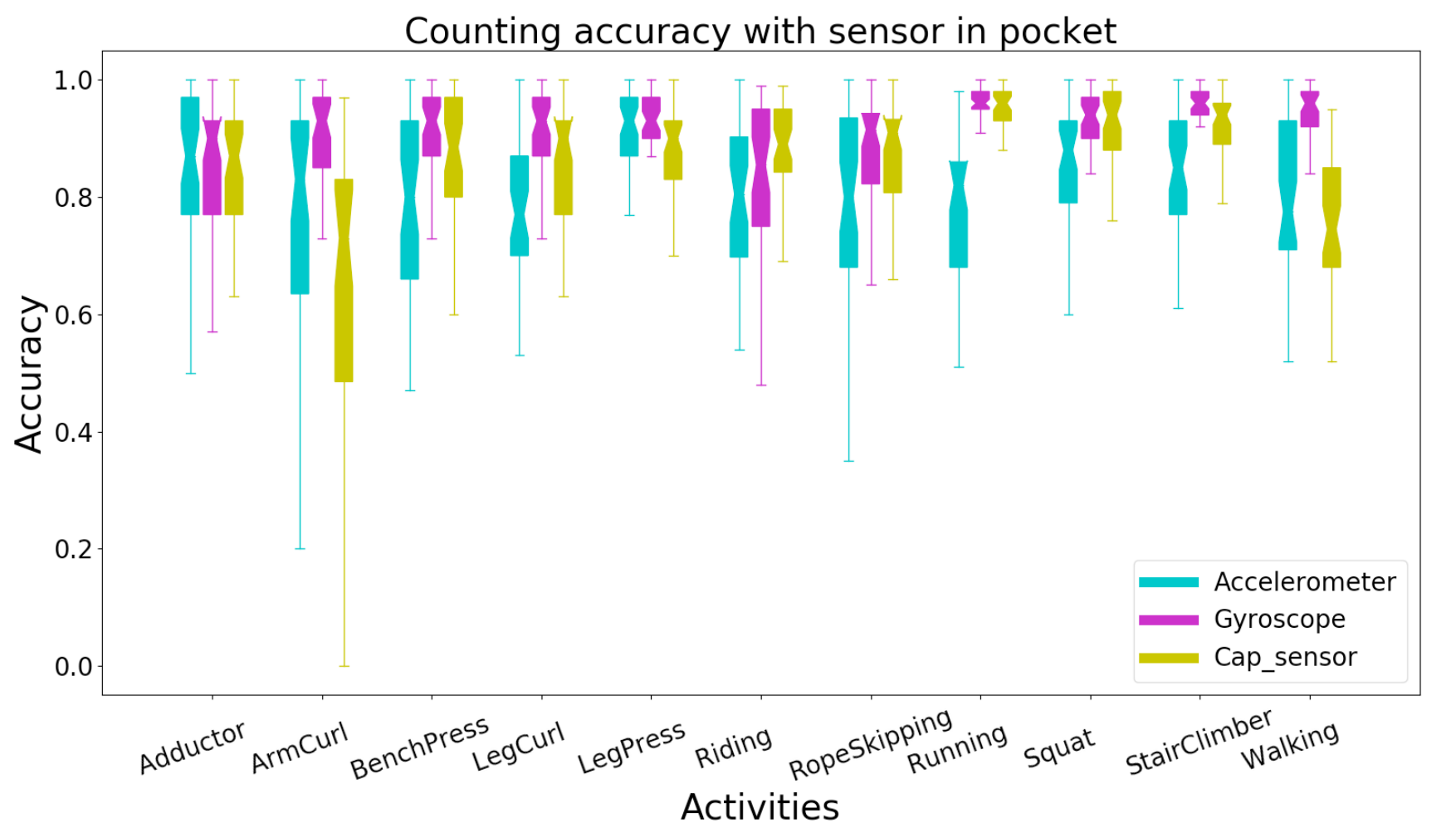}
\label{accuracy_pocket}
\end{subfigure}%
\centering

\caption{Counting accuracy from three sensing sources with three prototype deployments}
\label{Counting_three}
\end{figure}

\subsection{classification Result}
Figure \ref{Attention_CM} depicts the classification result. Using the hybrid model, we got the accuracy of gym workout recognition up to 80.1\%, 90.29\%, and 94.42\% with the combined sensing method when the prototype was in the pocket, on the calf and wrist, respectively, which outperforms the traditional classification methods like random forest and other typical deep models in human activity recognition like Resnet and DeepConvLSTM \cite{bian2022contribution}, as Table \ref{Deep_Resualt} lists. 
The macro F-score shows a similar trend among the three deployment positions. The depicted result is stable during several rounds of repetitive runs with less than 0.2\% accuracy deviation in each case.
In the single signal source test, the $IMU$ shows very near classification accuracy to the combined case (79.81, 88.94, 93.57 in percentage), meaning that it covers most of the recognition contribution. The $HBC$ shows weak classification ability compared to $IMU$ and brings around an averaged 1\% increase to the $IMU$ methods in the combined case, especially for activities when $IMU$ is relatively static, like the wrist-worn prototype for recognizing the Adductor workout, where the $HBC$ signal enhances the accuracy from 77\% to 80\%. Another noticeable recognition variation (wrist-worn case) when adding $HBC$ is the class of Walking, where the $IMU$ provides 94\% accuracy and $HBC$ drags it to 91\%, where the disappeared 3\% is mostly classified to the Null activity. However, in an averaged view regarding the Null class, the $HBC$ provides a 2\% increase to the $IMU$ solution. For the other ten classes in the wrist-worn case, the accuracy variations are very slight, resulting in a final contribution of only 0.85\% from $HBC$. In the leg-worn case, $HBC$ impressively improves the Walking recognition from 87\% to 93\%, mostly corrected from the Null class. For the other eleven classes, $HBC$ shows different supportive levels (from -1\% to 3\% variations when adding $HBC$ to $IMU$). And an overall 1.35\% positive regulation is averagely observed. In the pocket-worn case, the influence of adding $HBC$ to $IMU$ is more diverse, ranging from -5\% to 4\%, and an averaged positive influence of 0.2\%. Considering the accuracy deviation of several rounds of training results (0.2\% variation), pocket-worn $HBC$ shows no impact on the twelve activities recognition with a pocket-worn $IMU$. We suppose the reason behind this should be the weak coupling of the body surface and the sensing antenna of the prototype.


The weak contribution of $HBC$ to $IMU$ is as expected and reasonable. There are several reasons for it:
First, $IMU$ is sensitive even to subtle vibrations; thus, it still perceives the damped motion action even when it looks "static". In this case, one of the most important advantages of $HBC$, the cross-body part motion sensing, pitiably lost the chance of a sound demonstration. 
Second, the $HBC$ sourced signal is weak in robustness, and the signal pattern relies on the action scale related to the ground, which is highly subject-dependent. 
Third, the $IMU$ supplies six channels (3-axes accelerometer,3-axes gyroscope), and both the channel-wise temporal feature and the cross-channel spatial feature are extracted for representation, while the $HBC$ could only supply single-channel temporal features.

\begin{table*}[]
\centering
\begin{threeparttable}

\begin{tabular}{p{1.5cm} p{1.8cm} p{1.8cm} p{1.8cm} p{1.8cm} p{1.8cm}}
Position & Acc & Gyro & $HBC$ & $IMU$ & $HBC$ + $IMU$ \\ 
\hline
Pocket & 0.806$\pm$0.159 & 0.797$\pm$0.168 & \textbf {0.811}$\pm$0.168 & 0.809$\pm$0.152 &  \textbf {0.812}$\pm$0.156  \\
\hline
Leg & 0.801$\pm$0.158 & 0.788$\pm$0.181& \textbf {0.820}$\pm$0.190 & 0.802$\pm$0.160 &  \textbf {0.817}$\pm$0.162\\ 
\hline
Wrist & 0.752$\pm$0.223 & 0.739$\pm$0.219 &  \textbf {0.800}$\pm$0.217 & 0.756$\pm$0.190 &  \textbf {0.780}$\pm$0.204\\ 
\hline
\end{tabular}
\caption{Averaged counting accuracy with/without $HBC$ sensing modality}
\label{Counting_Cap_2}
\end{threeparttable}
\end{table*}

\begin{figure*}[]
\centering
\includegraphics[width=1.0\textwidth,height = 7.0cm]{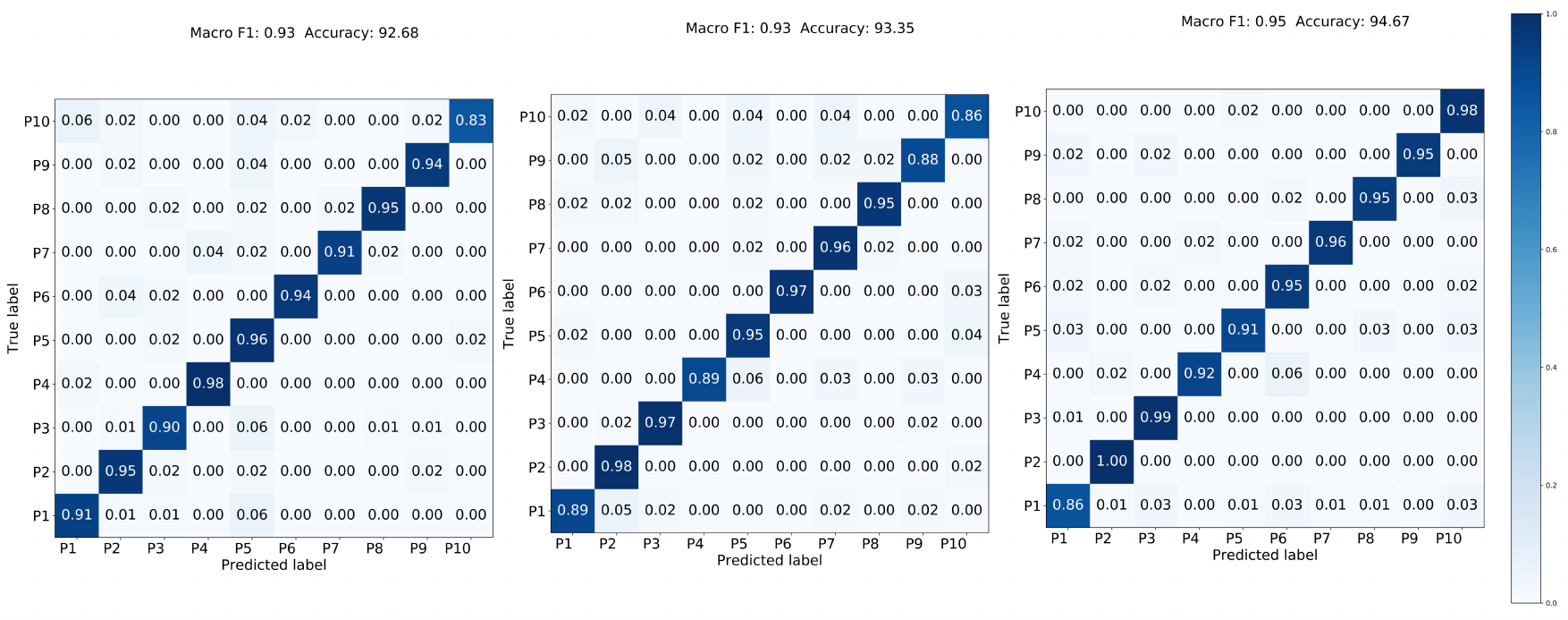}
\caption{User authentication with the prototype worn on wrist, leg, and pocket (left to right), using the combined signal of Running activity}
\label{SubjectClassificaition}
\end{figure*}

\subsection{Exercise Counting Exploration}
We performed counting exploration directly on the original data in a separate track without first segmenting it using the classification pipeline. This provided an upper bound on the possible counting performance.
Three sources of data for counting were utilized: accelerometer, gyroscope, and $HBC$. First, we combined the three axes to one magnitude so that each source provides one data sequence.
Then, we used Fourier Transform and Inverse Fourier Transform to smooth the data, removing the undesired high frequencies. Figure \ref{Counting_examples} gives a peak counting example of the Adductor workout from the first subject, where the wrist-worn $IMU$, in this case, almost lost the repetition cycles while the wrist-worn $HBC$ could still clearly capture them. To be noticed, a different cut-off frequency (5Hz) was given to the Fourier process for Running, Walking, Ropeskipping, and Riding, as they have a higher frequency than the other exercises (2.5Hz cut-off frequency) and are also highly recognizable by the classification model (e.g. 100\% group recognition accuracy to the other workouts with the wrist-worn combined configuration). Even in the worst classification case of $HBC$-pocket configuration, those four workouts are misclassified mostly among themselves and easily distinguishable against other workouts.



Finally, we detected the peaks of the smoothed signal using the PeakUtils \cite{PeakUtils} python package. Two parameters were set after grid searching to identify the peaks correctly. One is the threshold in relation to the highest value, and another is the minimum distance between two peaks.
We use accuracy(\begin{math} 1.0 - \frac{\abs{count_{detected} - count_{real}}}  { count_{real} }\end{math}) to present the counting performance. Figure \ref{Counting_three} uses Boxplot to show the counting result with signals from the accelerometer, gyroscope, and $HBC$ separately for each gym workout. In most cases, the capacitance sensing performs better than the accelerometer or gyroscope for counting, which was also represented in Table \ref{Counting_Cap_2} with the columns $Acc$, $Gyro$, $HBC$, where the counting accuracy is summarized regardless of exercise type.
Grouping the accelerometer and gyroscope by averaging the calculated count gives the $IMU$ result. Then, grouping the three by choosing the closest two and averaging them gives the $HBC$ + $IMU$ result, which improves the reliability of the fusion result and is better than averaging the three directly. In any case (single signal source or fused signal sources), $HBC$ supplies the first-class repetition counting result over $IMU$.



\subsection{user authentication Exploration}
Besides the workout recognition and counting evaluation with the designed prototype, we also exploited other potential functional capabilities, e.g. user authentication. Figure \ref{SubjectClassificaition} depicts the user authentication result among the ten participants using the combined signal of the Running workout (best authentication result among the eleven workouts). Considering the data volume used for training, we only applied the CNN block of Figure \ref{Network} followed directly by the classifier composed of dense and softmax layers.  As a result, we got an accuracy of 92\% - 95\% with the three prototype positions. Although the number of classified subjects is limited, this result is meaningful in the practical scenario, for example, for a family-shared gym tracking device.


\begin{figure}[!b]
\centering
\includegraphics[width=0.99\linewidth,height=9.0cm]{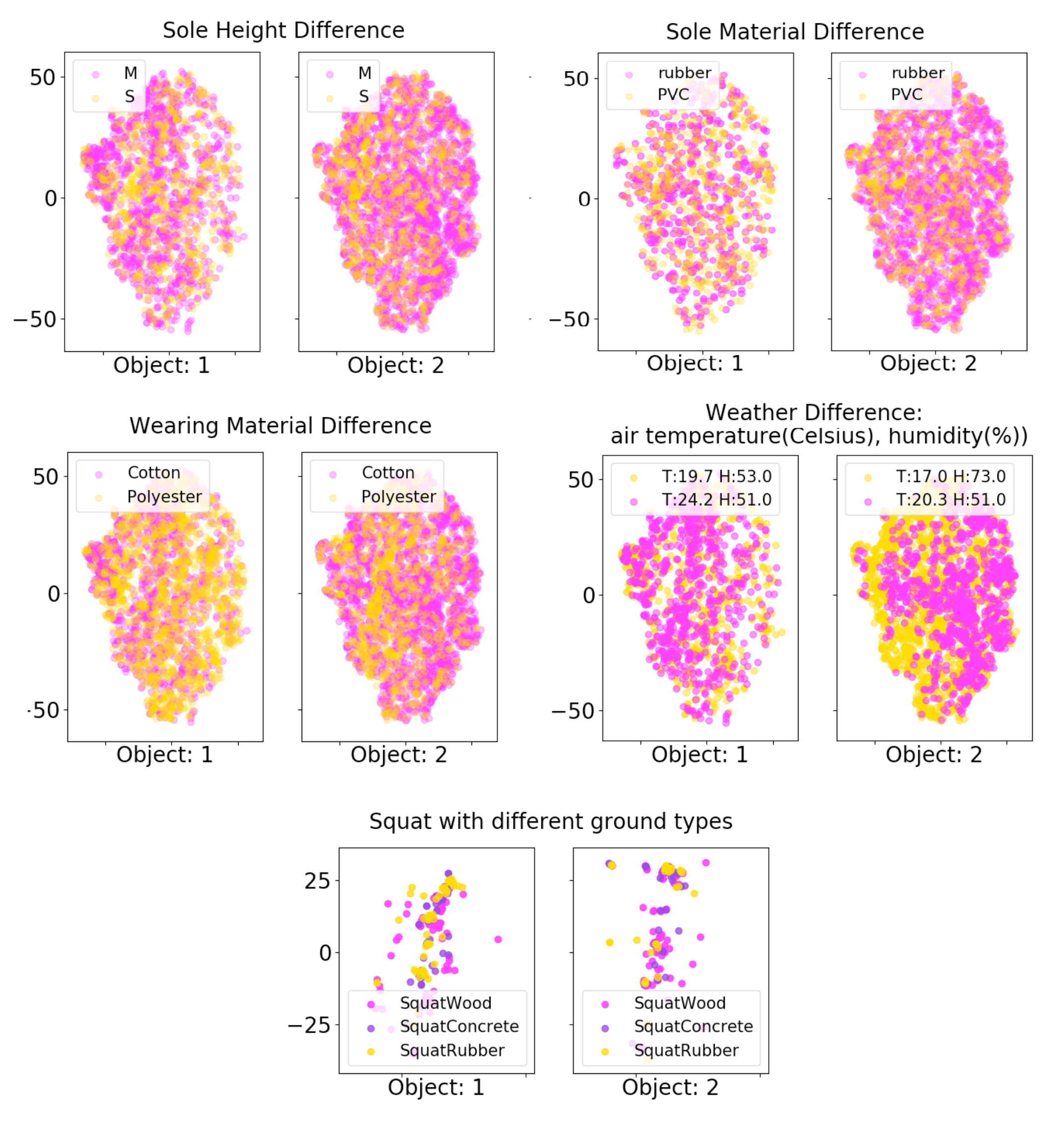}
\caption{t-SNE: Factors that may influence the robustness of $HBC$ based workout recognition}
\label{robustness}
\end{figure}

\subsection{Factors that impact variation of $HBC$}
This section analyzes the potential influence factors for the $HBC$ based sensing modality. It is clear that the wearing, especially the type and height of the sole, will affect the body capacitance \cite{jonassen1998human}. Since the designed prototype does not measure the body capacitance directly, instead, it continuously measures the body surface potential's variation caused by the variation of body capacitance. It is not clear if the wearing and other potential factors also influence the measured body surface potential variation (variation scale of body capacitance). Thus, we inspected five possible influence factors: Sole Height(M or S), Sole Material(PVC or rubber), Wearing(cotton or polyester), Weather Condition, and Ground Type(wood, concrete, or rubber). The configuration of wearing and the shoe is listed in Table \ref{Subjects_Configuration}. The weather condition was measured by HMP45D, offering the air temperature and the relative humidity. The ground type was verified by doing the Squat workout on three ground types. 
We studied the five factors within a single subject to avoid bias across subjects. Since data from a single subject is deficient for adopting the neural network classification method (easily resulting in overfitting), we use t-SNE here to roughly describe the feature distribution considering the above-listed five categories, as Figure \ref{robustness} (from two objects) depicted, which 
does not show an obvious separable distinction. From this point, we could preliminarily conclude that the considered factors do not influence the robustness of the classification and counting result. Our on-site observation also showed that the body motion itself dominates the body surface potential variation almost noncompetitive. 
The weather condition shows light separation on object 2's features distribution. However, it does not give the same result on the other subjects' data. For a substantial and solid influence factor demonstration, specific conditioned experiments would need to be designed, and we leave this research topic open for future exploration.

\section{Discussion}

$HBC$ sensing has its merit for wearable computing and depicts the potential to be a novel sensing modality for human activity recognition (HAR), for example, in gym workout recognition and repetition counting. This section discusses its advantages, limitations, future potential in the context of HAR, and a few considerations regarding practical implementation.

\subsection{Advantages}
Unlike $IMU$, which mainly captures localized motion, $HBC$ leverages the body’s electrostatic field, allowing it to sense movements beyond its attachment point. This property enhances activity recognition, particularly when the sensor is positioned on body parts that remain relatively static (e.g., wrist-worn devices detecting leg movements). As demonstrated in this paper, $HBC$ shows superior performance in workout repetition counting compared to $IMU$. The ability to detect cross-body part motion variations enables more precise tracking of repetitive movements, reducing common $IMU$ issues such as information loss during on-body motion transmission. Besides this, $HBC$-based sensing is cost-effective and energy-efficient, making it suitable for long-term wearable applications. The sensing front-end requires only a few discrete components, consuming power at the microwatt level, significantly lower than many $IMU$-based systems. Finally, $HBC$ signal processing is relatively lightweight, requiring only simple filtering and peak detection for repetition counting, making it ideal for low-power embedded systems.

\subsection{Limitations}
However, current studies on $HBC$ are still limited, and the preliminary outcomes (such as those concluded in this paper) show significant challenges to making $HBC$ a ubiquitous signal source. One challenge of $HBC$ is that $HBC$ is an interactive signal describing the electrical relationship between the body and the environment. This means $HBC$ varies under different environments \cite{bian2021systematic} (e.g., indoors and outdoors), making the usage of $HBS$ difficult when the robustness of the system is strictly required. Another challenge is the weak contribution in complex applications like activity recognition, as shown in this paper, that $IMU$ alone achieves near-identical accuracy to the combined $IMU$ and $HBC$ approach, indicating that $HBC$ does not significantly enhance overall recognition performance. Unlike $IMU$, which remains stable across different conditions, $HBC$-based measurements may exhibit non-linear fluctuations, potentially impacting accuracy in uncontrolled settings, and factors such as physical contact quality may also introduce inconsistencies in performance. While $HBC$ improves repetition counting, its contribution to workout classification is relatively weak when compared to $IMU$. 

\subsection{Future Exploration around $HBC$}
Considering the ubiquitous applications that can be boosted by $HBC$ and the current development and challenges of $HBC$ in wearable computing, we proposed two potential directions for future exploration (from the hardware side and algorithm side, respectively), aiming for a more accurate, robust, easy-of-use sensing with $HBC$.

\subsubsection{Active Shielding}

Considering the susceptibility of $HBC$ to environmental interference, active shielding could be an effective technique to minimize unwanted artifacts in sensor readings. By incorporating a shielding layer alongside the sensing electrode, external disturbances can be mitigated, leading to an improved signal-to-noise ratio (SNR) and ensuring that only the targeted region is detected. Several previous studies on capacitive sensing have already integrated active shielding mechanisms into their sensor front-end design. For instance, in \cite{takano2017non}, researchers developed a five-layer double-shield electrode to enhance resistance to external disturbances during non-contact electrocardiogram (ECG) and respiratory measurements. Their design positioned the sensing electrode as the first layer, a ground plane as the fifth, and introduced a driven shield at the third layer, which was connected to the output of a voltage follower. This configuration provided two key benefits: first, it minimized current leakage from the sensing electrode to the ground plane, thereby enhancing SNR, and second, it maintained high impedance at the sensing electrode, similar to a guard ring, to counteract external noise. The effectiveness of this approach was validated in an in-vehicle ECG system, where the double-shielded design significantly reduced vibratory and electromagnetic disturbances compared to conventional single-layer electrodes. Additionally, some commercial capacitive sensing chips already incorporate shielding techniques in their circuit architecture. A notable example is Texas Instruments’ FDC1004, which features an extra shielding pin to protect the capacitance input from electromagnetic interference (EMI) and to confine the sensing field to the intended area.

Considering the ubiquitous and complex interactions of $HBC$ with environmental factors, achieving precise capacitance measurements, a focused sensing area and a high SNR is crucial for ensuring robust and reliable performance. However, a comprehensive review of existing literature suggests that active shielding remains underutilized in many custom sensor front-end designs. Considering its proven benefits in capacitive sensing, along with its negligible impact on power consumption, size, and cost, it is likely that future $HBC$ sensing front-end designs will increasingly integrate active shielding for improved performance.

\subsubsection{Subject-dependent Continuous Learning}
Scaling $HBC$ sensing to large-scale, complex activity recognition remains challenging due to variability among users and environmental conditions. Traditional offline machine learning models typically rely on a static training dataset, meaning they cannot incorporate new knowledge after deployment. Consequently, over time, the accuracy of activity recognition may deteriorate as real-world conditions shift, such as changes in user behavior or environmental factors.

To ensure long-term reliability and adaptability, subject-dependent computing is a key research direction for making $HBC$-based applications practical in the real world. Online learning and incremental learning offer promising solutions by enabling models to continuously adapt to new data, refining their understanding of individual users and dynamic environments. Online learning frameworks generally consist of two primary components: model inference and model training, with retraining triggered whenever recognition accuracy declines.

For instance, \cite{disabato2022tiny} introduced a Tiny Machine Learning Concept Drift (TML-CD) approach that leverages deep learning feature extractors alongside a k-nearest neighbors (k-NN) classifier. Their system employed three adaptation strategies—passive update, active update, and hybrid update—to dynamically refine the k-NN model. The emergence of on-device training has further strengthened the feasibility of online learning, particularly for resource-constrained IoT and edge devices. To address hardware limitations, various approaches have been proposed, including reducing trainable parameters \cite{han2015deep, cai2019once} and minimizing activation memory requirements \cite{cai2020tinytl}. A recent algorithm-system co-design framework proposed in \cite{lin2022device} has demonstrated the feasibility of on-device lifelong learning with just 256KB of memory, allowing IoT devices to not only perform inference but also continuously learn from new data—a critical milestone for adaptive, subject-dependent learning on edge devices.

Currently, most $HBC$-based sensing studies rely on supervised learning models, where trained parameters remain fixed and do not account for potential shifts in data distribution caused by environmental changes or new users. Given the highly interactive nature of $HBC$ signals, integrating online learning with incremental updates would allow models to continuously refine parameters by leveraging unlabeled, incrementally collected user data. This capability would significantly enhance the adaptability and robustness of $HBC$-based HAR systems in real-world applications.

\subsection{Considerations for Practical Implementation}

The integration of $HBC$-based sensing into everyday wearable devices, such as smartwatches, presents several key challenges and considerations. The two primary aspects influencing real-world deployment are hardware miniaturization and gym tracking performance, as the outcomes studied in this work.

The current $HBC$ sensing front-end comprises multiple discrete components alongside an Analog-to-Digital Converter (ADC) for signal acquisition. While this configuration provides flexibility for experimental validation, it is not optimal for commercial, space-constrained wearables. A more practical alternative would be replacing the standalone ADC with a compact, low-power amplifier, reducing both circuit complexity and power consumption. Modern wearable microcontrollers (MCUs) and system-on-chip (SoC) architectures offer built-in low-noise amplifiers and high-resolution ADCs, which could eliminate the need for external ADC components. Furthermore, capacitive touch controllers, such as the FDC1004 from Texas Instruments, already feature shielded input channels to mitigate electromagnetic interference (EMI), which can be leveraged for $HBC$-based motion sensing in wearables.

Regarding gym tracking performance and feasibility, while $HBC$ contributes only marginally to activity classification accuracy, its effectiveness in repetition counting remains a notable advantage. $HBC$ could offer an alternative signal source that captures body-area electrostatic variations. This property enables accurate cycle detection, particularly for low-movement or stationary exercises, where $IMU$ signals alone may be insufficient (e.g., wrist-worn $IMU$ struggling with leg-dominant workouts).

We hope that the preliminary study presented in this work will encourage researchers to further investigations
considering the pervasive and promising usage scenarios backed by $HBC$ and enlighten researchers with a few study directions towards a more robust and feasible wearable motion sensing modality.

\section{Conclusion}

This work described a human body motion-tracking prototype composed of two sensing modalities, $IMU$ and $HBC$. $HBC$ features itself with low cost, low power consumption and enjoys cross-body-part attachment advantages over the $IMU$. We placed the prototypes in three body-area positions: in the pocket, on the calf, and on the wrist, and collected ten subjects' gym session data aiming to recognize the whole session of gym activity, and especially to evaluate the effectiveness of $HBC$ in wearable motion sensing with the $IMU$ solution as the background. A hybrid CNN-Dilated neural network empowered with the self-attention mechanism was designed for workout classification and a simple peak detection was applied for workout repetition counting. Results show that the $HBC$ sensing modality slightly boosts the traditional $IMU$ for activity recognition and obviously outperforms in repetition counting (obviously). Future work will be focused on the fundamental investigation of the body-area electrostatic field, especially on the influence factors, aiming to improve the understanding of the dynamic electrical relationship between the human body and its surroundings. We hope this work can reach the aim of providing a deeper impression of how $HBC$ can help with body motion detection in the wearable form and promoting more investigation and measures to bring this concept into practical applications.

\bibliographystyle{IEEEtran}
\bibliography{reference}{}

\end{document}